\documentclass{PoS}

\usepackage{graphicx}
\usepackage{feynarts}
\usepackage{amsmath}

\newcommand{\eg}{e.g.\ }

\newcommand{\ensuremathr}[1]{\ensuremath{\mathrm{#1}}}

\newcommand{\rd}{\ensuremathr{d}}

\newcommand{\rT}{\ensuremathr{T}}

\newcommand{\PZ}{\ensuremathr{Z}}
\newcommand{\PW}{\ensuremathr{W}}

\newcommand{\PWpm}{\ensuremathr{W^\pm}}

\newcommand{\Pu}{\ensuremathr{u}}

\newcommand{\Pd}{\ensuremathr{d}}

\newcommand{\Pp}{\ensuremathr{p}}

\newcommand{\Pj}{\ensuremath{j}}

\newcommand{\alphaw}{\ensuremath{\alpha_{\mathrm{w}}}}
\newcommand{\alphas}{\ensuremath{\alpha_{\mathrm{s}}}}

\newcommand{\GeV}{\ensuremathr{GeV}}
\newcommand{\TeV}{\ensuremathr{TeV}}
\newcommand{\nb}{\ensuremathr{nb}}

\newcommand{\order}[1]{\ensuremath{ {\mathcal{O}\left( #1 \right)} }}

\DeclareMathOperator{\myRe}{Re}

\newcommand{\proc}{\ensuremath{\Pp\Pp\longrightarrow \Pj\Pj+X}}

\newcommand{\sigtree}{\ensuremath{\sigma^0}}
\newcommand{\sigloop}{\ensuremath{\sigma^\mathrm{NLO}}}
\newcommand{\sigqcd}{\ensuremath{\sigma^0_\mathrm{QCD}}}

\newcommand{\deltree}{\ensuremath{\delta^\text{tree}_\mathrm{EW}}}
\newcommand{\delloop}{\ensuremath{\delta^\text{1-loop}_\mathrm{weak}}}
\newcommand{\delsum}{\ensuremath{\deltree{+}\delloop}}

\newcommand{\shat}{\ensuremath{\hat{s}}}
\newcommand{\that}{\ensuremath{\hat{t}}}
\newcommand{\uhat}{\ensuremath{\hat{u}}}

\newcommand{\kt}{\ensuremath{k_{\rT}}}

\newcommand{\ktl}{\ensuremath{k_{\rT,1}}} 

\newcommand{\mjj}{\ensuremath{M_{12}}}

\title{
Weak radiative corrections to dijet production\\
at the LHC
\thanks{Supported by the German Research Foundation (DFG) via grant DI 784/2-1.}
}

\ShortTitle{Weak radiative corrections to dijet production at the LHC}

\author{Stefan Dittmaier\\
        (Albert-Ludwigs-Universit\"at Freiburg)\\
        E-mail: \email{stefan.dittmaier@physik.uni-freiburg.de}}

\author{\speaker{Alexander Huss}\\
        (Albert-Ludwigs-Universit\"at Freiburg)\\
        E-mail: \email{alexander.huss@physik.uni-freiburg.de}}

\author{Christian Speckner\\
        (Albert-Ludwigs-Universit\"at Freiburg)\\
        E-mail: \email{christian.speckner@physik.uni-freiburg.de}}

\abstract{
We summarize the calculation of the weak corrections to dijet production at hadron colliders, comprising tree-level effects of $\order{\alphas\alpha,\;\alpha^2}$ and loop corrections of $\order{\alphas^2\alpha}$. 
Although suppressed by the small value of the coupling constant $\alpha$, the weak radiative corrections can become large in the high-energy domain due to the appearance of Sudakov-type and other high-energy logarithms. 
Generally the corrections to the transverse-momentum distributions are larger by approximately a factor of two compared to the corresponding reach in the invariant-mass distributions, because the invariant-mass distributions are not, unlike the $\kt$ distributions, dominated by the Sudakov regime at high scales.
The electroweak tree-level contributions are found to be of the same generic size as the loop corrections.
}

\FullConference{XXI International Workshop on Deep-Inelastic Scattering and Related Subjects\\
		 22-26 April, 2013\\
		 Marseilles, France}

\begin{document}

\section{Introduction}
\label{sec:introduction}

The inclusive dijet production $\Pp\Pp\to\Pj\Pj+X$ is an important process to test the Standard Model in the previously unexplored region that is now accessible at the LHC as well as in the search for physics beyond the Standard Model, see \eg Ref.~\cite{Harris2011}.
Furthermore, it delivers crucial constraints in the fit of the parton distribution functions (PDF), in particular for the gluon PDF at high momentum fraction $x$.

The next-to-leading order (NLO) QCD corrections have been calculated a long time ago~\cite{nloqcd}, and a substantial effort is currently put into the computation of the corrections at next-to-next-to-leading order (NNLO) in QCD, where the results for the purely gluonic channel have been presented in Ref.~\cite{Ridder:2013mf} recently.
Here we report on our calculation~\cite{Dittmaier:2012kx} of the purely weak radiative corrections of $\order{\alphas^2\alpha}$ to dijet production.
Corrections at this order have been previously calculated for the single-jet-inclusive cross section in Ref.~\cite{Moretti2006c}, and preliminary results to dijet production were shown in Ref.~\cite{Scharf2009}.

In spite of the suppression by the small value of the coupling constant $\alpha$, it is well known that the electroweak (EW) corrections can become large in the high-energy domain due to the appearance of Sudakov-type and other high-energy logarithms.
Considering that the data collected with the LHC running at the centre-of-mass (CM) energy of $\sqrt{s}=7~\TeV$ was already able to probe this high-energy domain of dijet invariant masses $\mjj$ and jet transverse momenta $\kt$ up to approximately $5~\TeV$ and $2~\TeV$, respectively, it is important to investigate the impact of these electroweak corrections.
Guided by the aforementioned logarithmic enhancements, we have restricted ourselves to the calculation of the purely weak radiative corrections at the order $\alphas^2\alpha$ in the first step, which will be denoted by $\alphas^2\alphaw$ in the following.
They form a well-defined gauge-invariant subset of the full EW corrections which can be supplemented by the remaining photonic QED corrections at a later time.

\section{Dijet production at hadron colliders}
\label{sec:dijet-production}

\begin{figure}[b]
  \centering
  {\unitlength=0.75bp
    \raisebox{30\unitlength}{\text{\footnotesize(a)}\quad
      $\left\lvert\vphantom{\rule{0mm}{35\unitlength}}\right.$}
    {\scriptsize
  \raisebox{-15\unitlength}[70\unitlength][0mm]{\input{feyndiags/TreeQCD}}
}
    \raisebox{30\unitlength}{$\left.\vphantom{\rule{0mm}{35\unitlength}}\right\rvert^2$}
    \qquad\quad
    \raisebox{30\unitlength}{\text{\footnotesize(b)}\quad
      $\left\lvert\vphantom{\rule{0mm}{35\unitlength}}\right.$}
    {\scriptsize
  \raisebox{-15\unitlength}[70\unitlength][0mm]{\input{feyndiags/TreeZ}}
}
    \raisebox{30\unitlength}{$+$}
    {\scriptsize
  \raisebox{-15\unitlength}[70\unitlength][0mm]{\input{feyndiags/TreeW}}
}
    \raisebox{30\unitlength}{$\left.\vphantom{\rule{0mm}{35\unitlength}}\right\rvert^2$}
    \\[1em]
    \raisebox{30\unitlength}{\text{\footnotesize(c)}\quad
      $2\,\myRe\,\left\lbrace\vphantom{\rule{0mm}{35\unitlength}}\right.$}
    {\scriptsize
  \raisebox{-15\unitlength}[70\unitlength][0mm]{\input{feyndiags/TreeQCD}}
}
    \raisebox{30\unitlength}{$\times$}
    \raisebox{30\unitlength}{$\left(\vphantom{\rule{0mm}{32\unitlength}}\right.$\hspace{-5\unitlength}}
    {\scriptsize
  \raisebox{-15\unitlength}[70\unitlength][0mm]{\input{feyndiags/TreeW}}
}
    \raisebox{30\unitlength}{\hspace{-5\unitlength}$\left.\vphantom{\rule{0mm}{32\unitlength}}\right)^*$}
    \raisebox{30\unitlength}{$\left.\vphantom{\rule{0mm}{35\unitlength}}\right\rbrace$}
  }
  \caption{The tree-level contributions to the process $\Pu\Pd\to\Pu\Pd$ of the orders (a)~$\alphas^2$, (b)~$\alpha^2$, and (c)~$\alphas\alpha$.}
  \label{fig:tree}
\end{figure}

When investigating the EW effects in dijet production one first has to note that already at leading order (LO) there are EW contributions in case of the four-quark processes given by the exchange of an electroweak gauge boson between the two quark lines.
This leads to the Born cross section not only consisting of the purely QCD contributions of $\order{\alphas^2}$, but also from interference and squared contributions of $\order{\alphas\alpha,\;\alpha^2}$.
The different diagrams and their respective contribution to the different orders in case of the subprocess $\Pu\Pd\to\Pu\Pd$ are shown in Fig.~\ref{fig:tree}.
Note that only the product between the $t$-channel and $u$-channel diagram gives a non-vanishing contribution to the interference term of $\order{\alphas\alpha}$ due to the colour structure.
In the LO cross section the photonic contributions are fully taken into account.

\begin{figure}
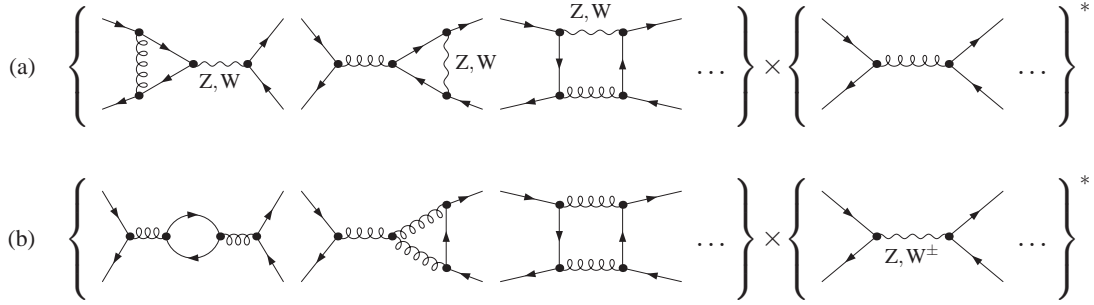

  \centering
  {\unitlength=0.75bp
  \begin{align*}
    \raisebox{30\unitlength}{\text{\footnotesize(a)}\quad
      $\left\lbrace\vphantom{\rule{0mm}{35\unitlength}}\right.$}
    {\scriptsize
  \raisebox{-15\unitlength}[70\unitlength][0mm]{\input{feyndiags/VSW}}
}
    \raisebox{30\unitlength}{$\ldots\left.\vphantom{\rule{0mm}{35\unitlength}}\right\rbrace\times$}
    &\raisebox{30\unitlength}{$\left\lbrace\vphantom{\rule{0mm}{35\unitlength}}\right.$}
    {\scriptsize
  \raisebox{-15\unitlength}[70\unitlength][0mm]{\input{feyndiags/BS}}
}
    \raisebox{30\unitlength}{$\ldots\left.\vphantom{\rule{0mm}{35\unitlength}}\right\rbrace^*$}
    \\[0.5em]
    \raisebox{30\unitlength}{\text{\footnotesize(b)}\quad
      $\left\lbrace\vphantom{\rule{0mm}{35\unitlength}}\right.$}
    {\scriptsize
  \raisebox{-15\unitlength}[70\unitlength][0mm]{\input{feyndiags/VSS}}
}
    \raisebox{30\unitlength}{$\ldots\left.\vphantom{\rule{0mm}{35\unitlength}}\right\rbrace\times$}
    &\raisebox{30\unitlength}{$\left\lbrace\vphantom{\rule{0mm}{35\unitlength}}\right.$}
    {\scriptsize
  \raisebox{-15\unitlength}[70\unitlength][0mm]{\input{feyndiags/BW}}
}
    \raisebox{30\unitlength}{$\ldots\left.\vphantom{\rule{0mm}{35\unitlength}}\right\rbrace^*$}
  \end{align*}}
\vspace{-25\unitlength}
  \caption{The virtual corrections of $\order{\alphas^2\alphaw}$ illustrated by terms of some typical interferences.}
  \label{fig:virt}
\end{figure}

At NLO we restrict our calculation to the purely weak corrections at the order $\alphas^2\alphaw$ with a selection of diagrams for the virtual corrections shown in Fig.~\ref{fig:virt}.
Contributions at this order can be obtained by considering weak $\order{\alphaw}$ corrections to the Born QCD cross section ($\order{\alphas^2}$) or by considering QCD $\order{\alphas}$ corrections to the LO interference terms ($\order{\alphas\alphaw}$). 
A strict separation of the corrections is not possible, owing to the appearance of diagrams of the type such as the third one-loop diagram in Fig.~\ref{fig:virt}~(a), which could be attributed to both.
Instead, one has to consistently take into account all corrections defined by the order in perturbation theory.
A more extensive discussion of the calculational details can be found in Ref.~\cite{Dittmaier:2012kx}.

\section{Numerical results}
\label{sec:results}

We define a dijet event by requiring at least two jets with a transverse momentum $\kt > 25~\GeV$ each and a rapidity $y$ with $\lvert y \rvert < 2.5$, where we employ the anti-$k_\rT$ algorithm with the angular separation parameter of $R=0.6$ for the jet definition.
Further details on the numerical input can be found in Ref.~\cite{Dittmaier:2012kx}.
The NLO correction relative to the Born cross section $\sigtree$ is defined via $\sigloop=\sigtree\times(1+\delloop)$.
In order to quantify the impact of the LO EW contributions of $\order{\alphas\alpha,\;\alpha^2}$ which are omitted in purely QCD predictions, we further introduce a relative correction factor $\deltree$ with respect to the Born QCD cross section, $\sigtree=\sigqcd\times(1+\deltree)$.

\begin{figure}
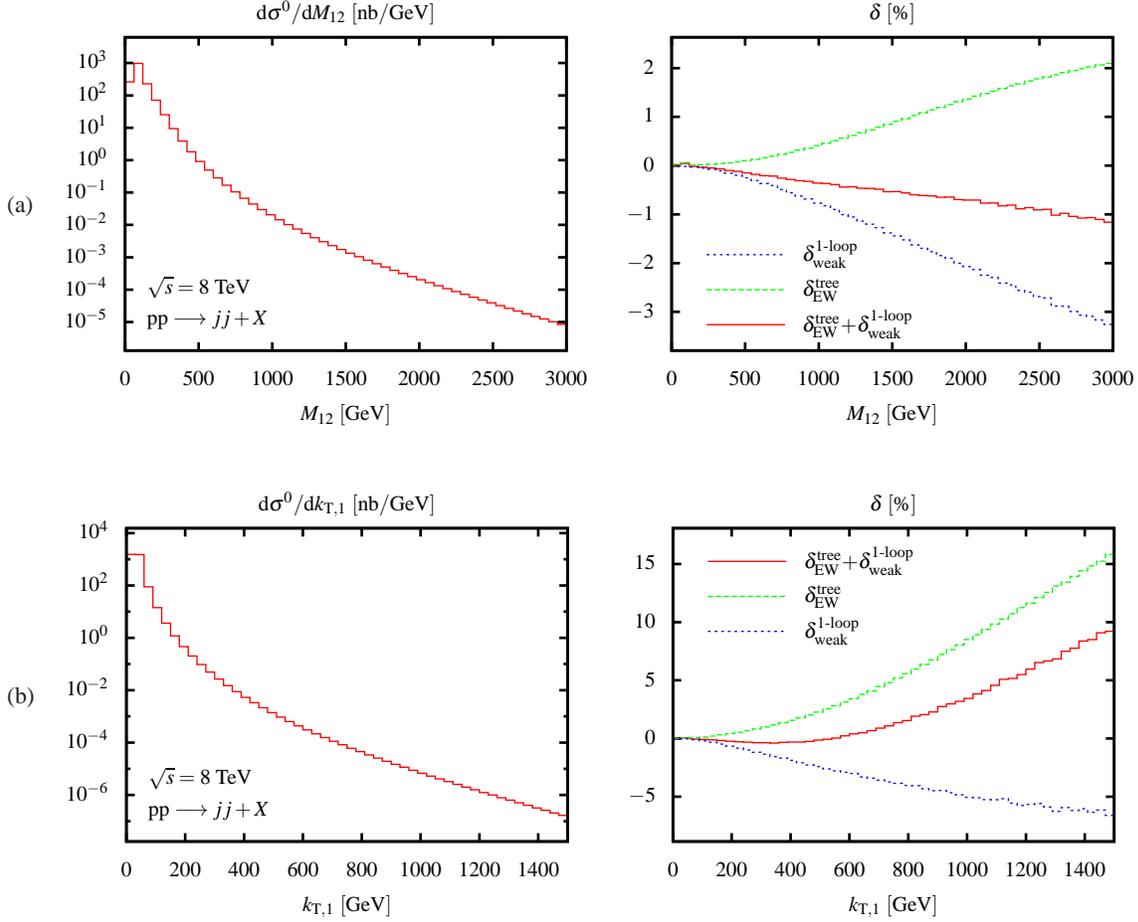

  \centering
  \scriptsize
  \raisebox{3cm}{\footnotesize(a)}~\input{plots/cdj8dist_MJJ.tex}\\
  \raisebox{3cm}{\footnotesize(b)}~\input{plots/cdj8dist_KT1.tex}
  \caption{Differential distributions with respect to (a)~the 
    dijet invariant mass $\mjj$ and (b)~the transverse momentum of the
    leading jet $\ktl$ at the LHC for a CM energy of 8~\TeV.
    Left: absolute predictions; right: relative contributions $\delta$
    (taken from Ref.~\cite{Dittmaier:2012kx}).}
  \label{fig:lhc8}
\end{figure}

The results for the LHC with the CM energy of $\sqrt{s}=8~\TeV$ are shown in Figs.~\ref{fig:lhc8}~(a,b) for the differential distributions with respect to the dijet invariant mass $\mjj$ and the transverse momentum of the leading jet, $\ktl$, respectively.
The weak radiative corrections show the typical behaviour expected from the Sudakov-type logarithms which are negative throughout and increase in magnitude at higher scales.
However, they turn out to be only of moderate size in case of the $\mjj$ distribution reaching approximately $-3\%$ for an invariant mass of $\mjj=2~\TeV$.
This can be understood by the fact that the high-$\mjj$ tail of the distribution is not dominated by the Sudakov regime where all scales (Mandelstam variables $\shat$, $\that$, $\uhat$) are simultaneously required to be much larger than the gauge-boson mass ($\shat,\,\lvert\that\rvert,\,\lvert\uhat\rvert \gg M_\PW^2$), but instead are dominated by the Regge (forward) region where $\shat$ is large but $\lvert\that\rvert$ or $\lvert\uhat\rvert$ remain small.
In case of the transverse-momentum distribution, on the other hand, the high-$\ktl$ domain probes the Sudakov-regime, and we observe larger NLO weak corrections, reaching around $-6\%$ for leading-jet transverse momenta of $\ktl=1.5~\TeV$.
The tree-level EW contributions are similar in size, but opposite in sign, leading to significant cancellations in the sum.
The rise of $\deltree$ with higher scales can be understood by inspecting the parton luminosities:
At lower values of $\mjj$ and $\ktl$ the cross section is dominated by the gluon-induced processes which do not contribute to the LO EW cross section.
The only non-vanishing contribution to $\deltree$ comes from the four-quark processes which gain in importance for higher scales, in contrast to the gluon-induced processes which become more and more suppressed due to the rapidly decreasing gluon luminosity.
In order to explain the larger corrections observed in the $\ktl$ distribution compared to the $\mjj$ distribution one needs to inspect the dominant contribution to $\deltree$ coming from the $\order{\alphas\alpha}$ interference terms of the valence quark--quark scattering: $q_1q_2\to q_1q_2,\;q_i\in\lbrace\Pu,\Pd\rbrace$. 
In the case for the subprocess $\Pu\Pd\to\Pu\Pd$ the different contributions to the LO cross section are given in Fig.~\ref{fig:tree}.
Owing to the colour structure, only the interferences between $t$- and $u$-channel diagrams deliver a non-vanishing contribution and lead to the observed effect that $\deltree$ is larger in the central region.

\begin{figure}
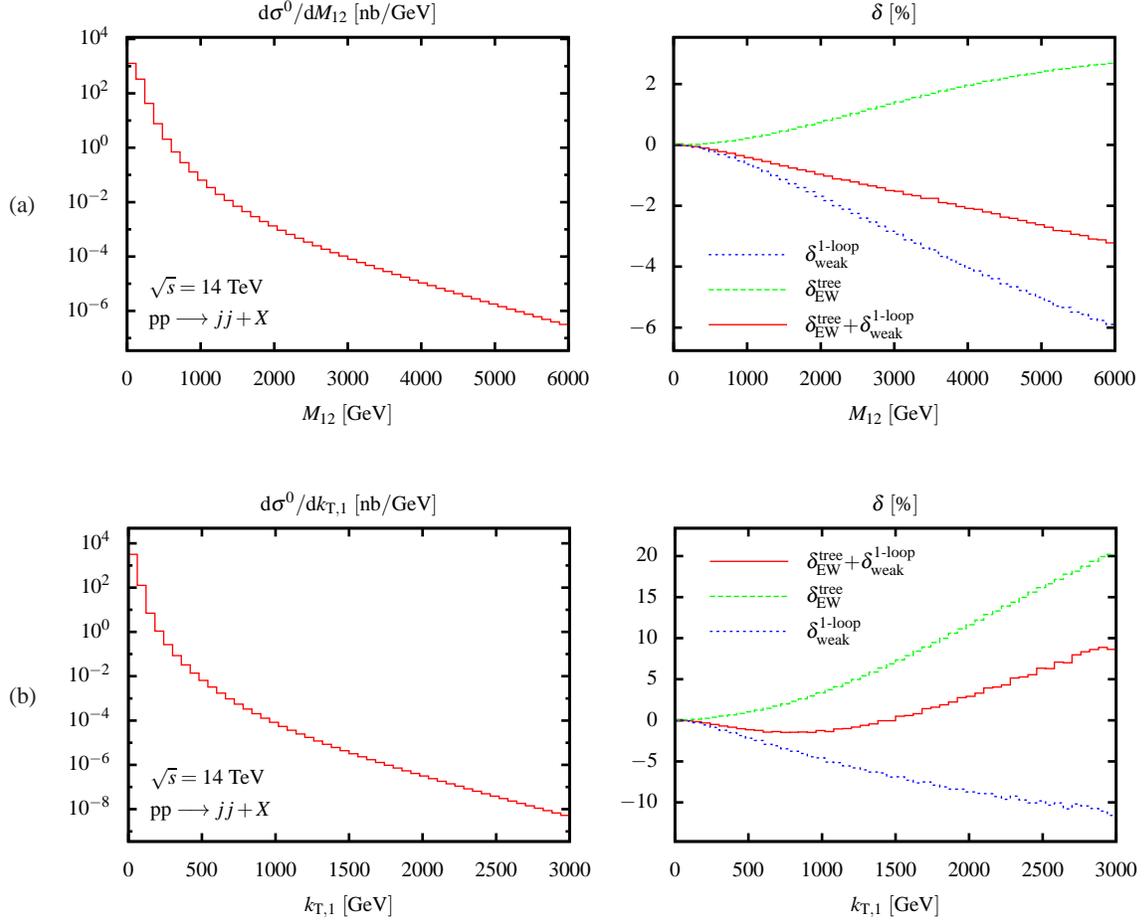

  \centering
  \scriptsize
  \raisebox{3cm}{\footnotesize(a)}~\input{plots/cdj14dist_MJJ.tex}\\
  \raisebox{3cm}{\footnotesize(b)}~\input{plots/cdj14dist_KT1.tex}
  \caption{Differential distributions with respect to (a)~the 
    dijet invariant mass $\mjj$ and (b)~the transverse momentum of the
    leading jet $\ktl$ at the LHC for a CM energy of 14~\TeV.
    Left: absolute predictions; right: relative contributions $\delta$
    (taken from Ref.~\cite{Dittmaier:2012kx}).}
  \label{fig:lhc14}
\end{figure}

Figure~\ref{fig:lhc14} shows the respective results for the LHC running at a CM energy of $\sqrt{s}=14~\TeV$, which exhibits an over-all behaviour similar to the $8~\TeV$ setup.
Owing to the deeper reach into the high-energy domain we observe larger loop corrections which amount to approximately $-6\%$ for an invariant mass of $\mjj=6~\TeV$ and $-11\%$ for leading-jet transverse momenta of $\ktl=3~\TeV$.
Although the tree-level EW contributions show a weaker dependence on the collider energy compared to the weak loop corrections, they are still similar in size leading to large cancellations in the sum.
The weak radiative corrections to the $\ktl$ distribution are again larger by almost a factor of two compared to the corresponding reach in the $\mjj$ distribution, in consequence of the fact that the high-$\ktl$ tail is, unlike the $\mjj$ distribution, dominated by the Sudakov regime.


\begin{thebibliography}{99}

\bibitem{Harris2011}
  R.~M.~Harris and K.~Kousouris,
  Int.\ J.\ Mod.\ Phys.\ A {\bf 26} (2011) 5005
  [arXiv:1110.5302 [hep-ex]].

\bibitem{nloqcd}
  S.~D.~Ellis, Z.~Kunszt and D.~E.~Soper,
  Phys.\ Rev.\ Lett.\  {\bf 64} (1990) 2121;
  Phys.\ Rev.\ Lett.\  {\bf 69} (1992) 1496;
  W.~T.~Giele, E.~W.~N.~Glover and D.~A.~Kosower,
  Phys.\ Rev.\ Lett.\  {\bf 73} (1994) 2019
  [hep-ph/9403347].

\bibitem{Ridder:2013mf}
  A.~G.~-D.~Ridder, T.~Gehrmann, E.~W.~N.~Glover and J.~Pires,
  arXiv:1301.7310 [hep-ph].

\bibitem{Dittmaier:2012kx}
  S.~Dittmaier, A.~Huss and C.~Speckner,
  JHEP {\bf 1211} (2012) 095
  [arXiv:1210.0438 [hep-ph]].

\bibitem{Moretti2006c}
  S.~Moretti, M.~R.~Nolten and D.~A.~Ross,
  Nucl.\ Phys.\ B {\bf 759} (2006) 50
  [hep-ph/0606201].

\bibitem{Scharf2009}
  A.~Scharf,
  arXiv:0910.0223 [hep-ph].

\end{thebibliography}
\end{document}